\documentclass[letterpaper, 10 pt, conference]{ieeeconf}  

\IEEEoverridecommandlockouts                              

\usepackage{amsmath} 
\usepackage{amssymb}  
\usepackage{tabularx}
\usepackage{subcaption}
\usepackage{booktabs}
\usepackage{graphicx}

\title{\LARGE \bf
Mixed Reality Tele-Ultrasound over 750 km: A Feasibility Study
}

\author{Ryan Yeung$^{1}$, David G. Black$^{2}$, 
Patrick Boyan Chen$^{2}$, 
Victoria Lessoway$^{3}$, 
Janice Reid$^{3}$, \\
Sergio Rangel-Suarez$^{4}$, 
Silvia D. Chang$^{4}$, 
Septimiu E. Salcudean$^{2}$
\thanks{*This work was supported in part by the National Science and Engineering Research Council of Canada (NSERC); in part by MITACS; in part by Rogers; and in part by the C.A. Laszlo Chair held by Prof. Salcudean.}
\thanks{$^{1}$Ryan Yeung is with the School of Biomedical Engineering,
        The University of British Columbia, Vancouver, BC V6T 1Z4, Canada
        {\tt\small ryeung1@student.ubc.ca}}%
\thanks{$^{2}$David G. Black, Patrick Boyan Chen and Septimiu E. Salcudean are with the Department of Electrical and Computer Engineering, The University of British Columbia, Vancouver, BC V6T 1Z4, Canada
        {\tt\small dgblack@ece.ubc.ca; patricchen2002@gmail.com; tims@ece.ubc.ca}}%
\thanks{$^{3}$Victoria Lessoway and Janice Reid are with The University of British Columbia, Vancouver, BC V6T 1Z4, Canada
        {\tt\small vickielessoway@shaw.ca; janreid@telus.net}}%
\thanks{$^{4}$Sergio Rangel-Suarez and Silvia D. Chang are with the Department of Radiology, The University of British Columbia, Vancouver, BC V6T 1Z4, Canada
        {\tt\small sergiocics@hotmail.com; Silvia.Chang@vch.ca}}%
}

\begin{document}

\maketitle
\thispagestyle{empty}
\pagestyle{empty}

\begin{abstract}

To address the lack of access to ultrasound in remote communities, previous work introduced human teleoperation, a mixed reality and haptics-based tele-ultrasound system. In this approach, a novice takes the role of a cognitive robot controlled remotely by an expert through mixed reality. In this manuscript we summarize new developments to this system and describe a feasibility study assessing it’s use for long-distance remote abdominal ultrasound examinations. To provide simple but effective haptic feedback, we used an ellipsoid model of the patient with its parameters calibrated using our system's position and force sensors. We tested the system in Skidegate, Haida Gwaii, Canada, with the experts positioned 754~km away in Vancouver, Canada. We performed 11 total scans with 10 novices and 2 sonographers. The sonographers were tasked with acquiring 5 target images in the epigastric region. The image acquisition quality was assessed by 2 radiologists. We collected alignment data and the novices completed task load and usability questionnaires. Both the novices and sonographers provided written and verbal feedback to inform future design iterations. 92\% of the acquired images had sufficient quality for interpretation by both radiologists. The mean task load reported by the novices was below reference values reported in literature and the usability was unanimously positive. No correlation was found between image quality and the follower’s alignment error with the virtual transducer. Overall, we show that human teleoperation enables sonographers to perform remote abdominal ultrasound imaging with high performance, even across large distances and with novice followers. Future work will compare human teleoperation to conventional, robotic and tele-mentored ultrasound.

\end{abstract}

\section{Introduction}
\subsection{Background}
Ultrasound (US) is a medical imaging modality widely used for the diagnosis of various diseases by allowing visualization of soft tissue structures. It offers many benefits due to its non-invasiveness, low-cost, and portability \cite{Law:2011}. US image acquisition is a technical skill performed by trained sonographers. Many remote communities lack individuals with this expertise. A qualitative study highlighted geographical isolation as a central barrier to US imaging for remote, Indigenous communities in Canada. Individuals in these communities must either wait for an itinerant sonographer who comes once per month, or travel up to 1040 km to a larger city for access to US imaging \cite{Adams:2021}. For instance, inhabitants of Skidegate, a small community on the islands of Haida Gwaii, BC, Canada, must take an 8 hour ferry to Prince Rupert where the closest major hospital is located to obtain a US exam. This travel can be a significant burden and cancelled ferries can lead to delayed diagnosis.

Tele-ultrasound (tele-US) aims to solve this problem by enabling an expert to remotely guide and perform US imaging. Robotic US is a form of tele-US in which an expert teleoperates a robotic arm to acquire the US images \cite{Jiang:2023}. Robotic US gives sonographers precise and responsive control and has been demonstrated in multiple clinical studies \cite{Delgorge:2005,Mathiassen:2016,Duan:2021,Adams:2022}. Yet, there has been limited commercial success which can likely be attributed to the high cost, large footprint, and complex maintenance, set-up and operation of a telerobotic system.

Tele-mentored US is another form of tele-US which uses video conferencing to enable a sonographer to guide US imaging performed by less qualified personnel \cite{Britton:2019}. Tele-mentored US is commonly used together with portable point of care US (POCUS) devices such as the Butterfly iQ3 (Butterfly Network, Burlington, MA) or Clarius C3HD3 (Clarius Mobile Health, Vancouver, BC). This makes the solution cheaper and easier to set up compared to robotic US. However, tele-mentored US does not allow precise control over the remote transducer so some expertise at the patient side is required. Additionally, there are few studies that demonstrate the efficacy of tele-mentored US \cite{Duarte:2022}.

Recently, there has been increased interest and advancements towards autonomous robotic US, largely driven by advancements in machine learning methods \cite{Su:2024,Ning:2023,Raina:2023}. A review of the challenges and perspectives related to machine learning in robotic US is presented in \cite{Bi:2024}. Benefits of autonomous robotic US compared to traditional or tele-US include improved reproducibility and a lack of reliance on experienced sonographers. While machine learning holds much promise for advancing autonomous robotic US, there still exists challenges regarding its ability to interpret, reason and make decisions during image acquisition. Furthermore, concerns regarding safe human-robot interaction and ethics limit its translational feasibility.

\subsection{Human Teleoperation}
To address the need for a low-cost, flexible, and precise tele-US solution, mixed reality (MR) ``Human Teleoperation" was introduced in \cite{Black:2023} and expanded in \cite{Black:2024}. In healthcare, MR has been used for surgical training, 3D anatomical visualization \cite{Sanchez-Margallo:2021}, and improving navigation in ultrasound-guided needle biopsies \cite{Groves:2022}. In human teleoperation, rather than rendering static overlay images or pointers, MR is used to render a dynamic virtual guiding US transducer, controlled in real time by a remote expert. The novice (``follower") follows the virtual transducer, acting as a flexible, cognitive robot. This approach offers greater precision than tele-mentored US while being more accessible and portable than robotic US.

The feasibility of using human teleoperation for tele-US remains unexplored. Previous work \cite{Black:2023:IJCARS} did not replicate the small, precise motions required for US, nor did it account for communication latency over large distances \cite{Black:2024:TMRB}. Additionally, understanding the impressions from individuals living in remote communities is imperative for guiding future translation. To address these gaps, we conducted 11 remote abdominal US scans using human teleoperation to answer the following questions:

\begin{enumerate}
    \item To what extent can inexperienced individuals from remote communities use the system as a follower?
    \item To what extent can interpretable US images be acquired in a truly remote setting using the system?
\end{enumerate}

By exploring these questions, our primary objective was to assess the feasibility of using human teleoperation for remote abdominal US imaging. We also gathered valuable feedback from both the followers and experts. In this paper, we first summarize the human teleoperation system, including a new and intuitive method for fitting an ellipsoid to the patient's torso for delay-robust force feedback to the sonographer. We then describe the feasibility tests and present results on usability and task load, follower alignment error, and US image quality.

\section{Methods}
\subsection{System Design}
\begin{figure}[t]
    \centering
    \includegraphics[width=\columnwidth]{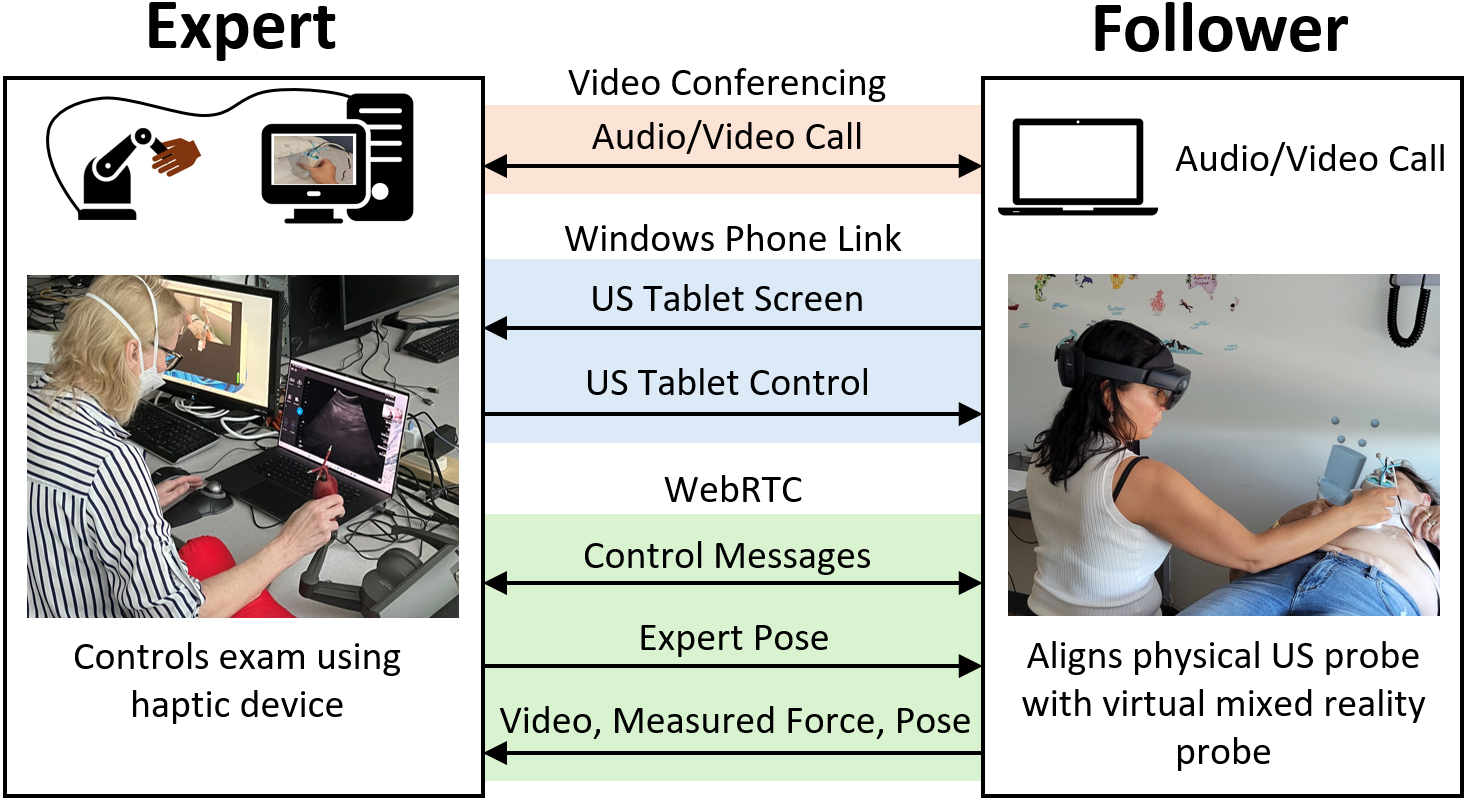}
    \caption{Diagram of human teleoperation system, showing the method and direction of communication.}
    \label{fig:humanTeleop}
\end{figure}

\begin{figure}[t]
    \centering
    \includegraphics[width=\columnwidth]{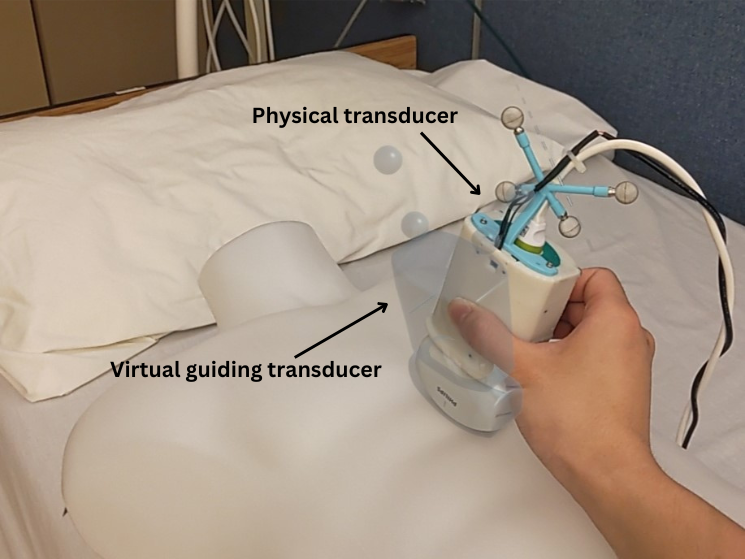}
    \caption{Follower's point of view. A virtual holographic guiding transducer is rendered through the Microsoft HoloLens 2.}
    \label{fig:followerUI}
\end{figure}

\begin{figure}[t]
    \centering
    \includegraphics[width=\columnwidth]{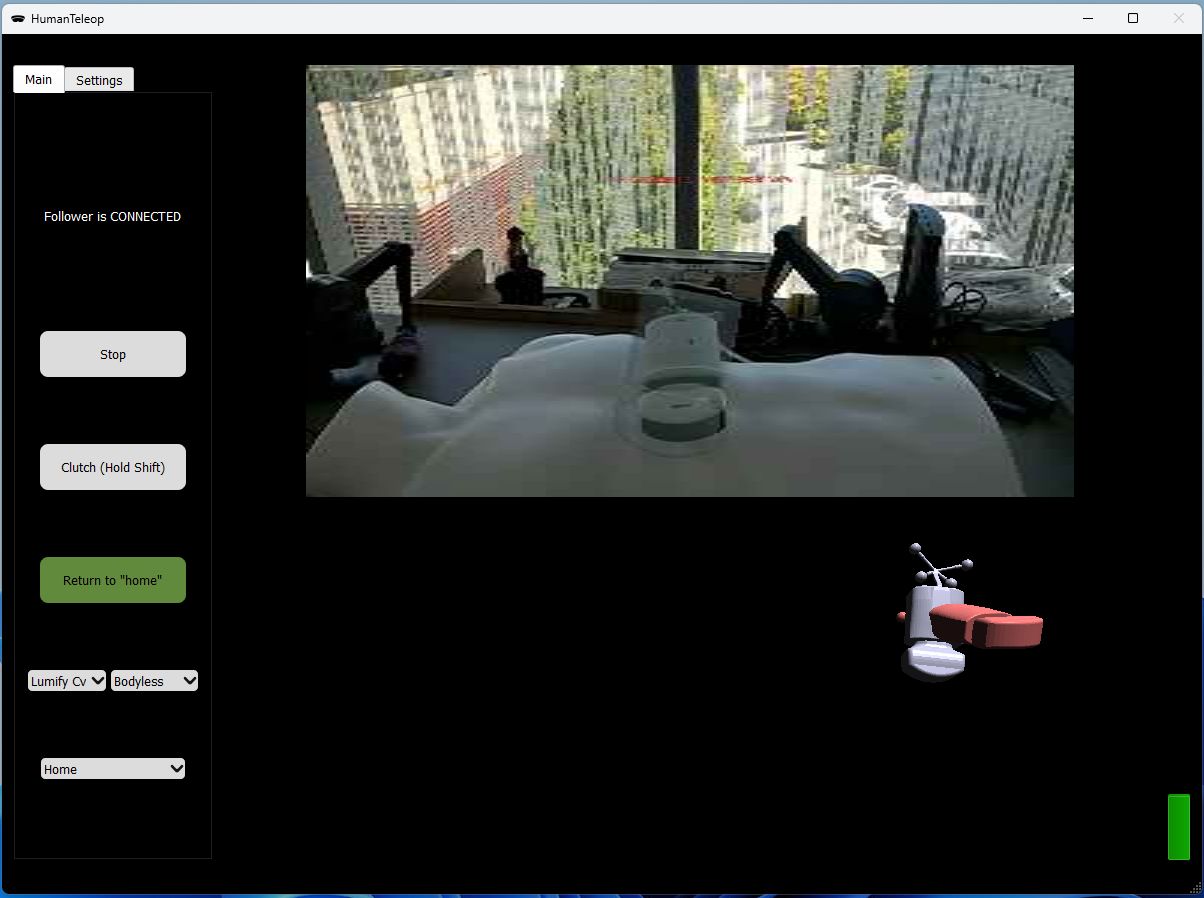}
    \caption{Expert's user interface which shows the point-of-view video from the follower. A secondary screen (not shown) was used to show the US image feed.}
    \label{fig:expertUI}
\end{figure}

In this study, we focus on the use of human teleoperation for tele-US. The system details are provided in references \cite{Black:2023, Black:2024:TMRB, Black:2024:TIM, Black:2024:DMFS} and are summarized here and in Fig. \ref{fig:humanTeleop} for completeness. For this application, a novice follower situated with the patient wears the Microsoft HoloLens 2 (Microsoft, Redmond, WA) MR headset which displays a virtual guiding US transducer in their environment. The follower superimposes the real US transducer on the virtual transducer and maintains alignment as the virtual transducer is moved. This is illustrated in Fig. \ref{fig:followerUI}. The pose of the virtual transducer is controlled by an expert sonographer located remotely (eg. at a medical center) using the Touch X haptic device (3D Systems, Rock Hill, SC) while seeing the real-time feed of the US image and point-of-view video from the follower. The expert's user interface is shown in Fig. \ref{fig:expertUI}. The haptic device is a desktop serial manipulator with 6-degree-of-freedom (DOF) input and can apply 3-DOF forces on its handle. As the sonographer moves the handle of the manipulator, its pose is measured and sent to the HoloLens 2 to be replicated by the virtual transducer. Attached to this handle is a 3D printed transducer-shaped model so the sonographer can feel like they are holding a real transducer. The haptic device also renders forces to the sonographer to mimic the sensation of a real US scan.

The communication between the follower side and expert side of the system was built using the Web Real Time Communication (WebRTC) peer-to-peer architecture. This is described in detail and evaluated in \cite{Black:2024:TMRB, Black:2024}. The measured transducer force and pose, and the point-of-view video feed of the follower are sent to the expert while the expert's haptic device pose is transmitted to the follower. The US image feed is shared over the Windows Phone Link app which allows the expert to view the US image and have complete real-time control of all the functions available on the Philips Lumify Ultrasound app, such as controlling the image parameters. The follower and expert communicate verbally over an external video call.

The pose of the US transducer is measured by attaching a known arrangement of infrared (IR) reflective markers and using the IR camera and time-of-flight depth camera on the HoloLens 2 \cite{Black:2024:TIM}. To measure the forces applied on the US transducer, differential magnetic force sensing is used which requires attaching a low-profile shell around the US transducer \cite{Black:2024:DMFS}.

\subsection{Patient Model for Haptic Feedback}
To provide accurate haptic feedback, the patient's abdominal surface is modeled as an ellipsoid, with forces computed based on its interaction with the virtual transducer. Points for the ellipsoid are collected by pressing the US transducer on four specific spots on the abdomen and bed surface. When transducer forces exceed a threshold due to patient contact, the tracked pose determines the tip position, enabling intuitive point collection at the scan's start. Due to large delays from communication and follower lag, direct force feedback was found to be impractical. The ellipsoid method, however, offered the expert a stable and relatively accurate haptic representation of the patient on which to move and rest their hand while guiding the scan. 

Several assumptions allow us to minimize the number of points required to fit the ellipsoid. Firstly, we assume the ellipsoid axes are aligned with the patient's longitudinal, sagittal and frontal axes respectively. This assumption holds as long as the patient is lying in supine position which is the case for abdominal scans in the epigastric region. Second, we set the longitudinal semi-axis (which we denote $c$) of the ellipsoid to a very large value since the accuracy of this dimension (along the height of the patient) is not necessary. Lastly, we assume the ellipsoid is tangent to the bed surface. With these assumptions we can determine the ellipsoid center $(z_c,y_c,x_c)$ and remaining two semi-axes lengths ($a$ and $b$) with the following four points:
\begin{enumerate}
    \item Patient's xiphoid process just below the sternum $(z_1,y_1,x_1)$
    \item Patient's extreme left $(z_2,y_2,x_2)$
    \item Patient's extreme right $(z_3,y_3,x_3)$
    \item Bed level $(z_4,y_4,x_4)$
\end{enumerate}

\begin{figure}[t]
 \centering
 \includegraphics[width=0.9\columnwidth]{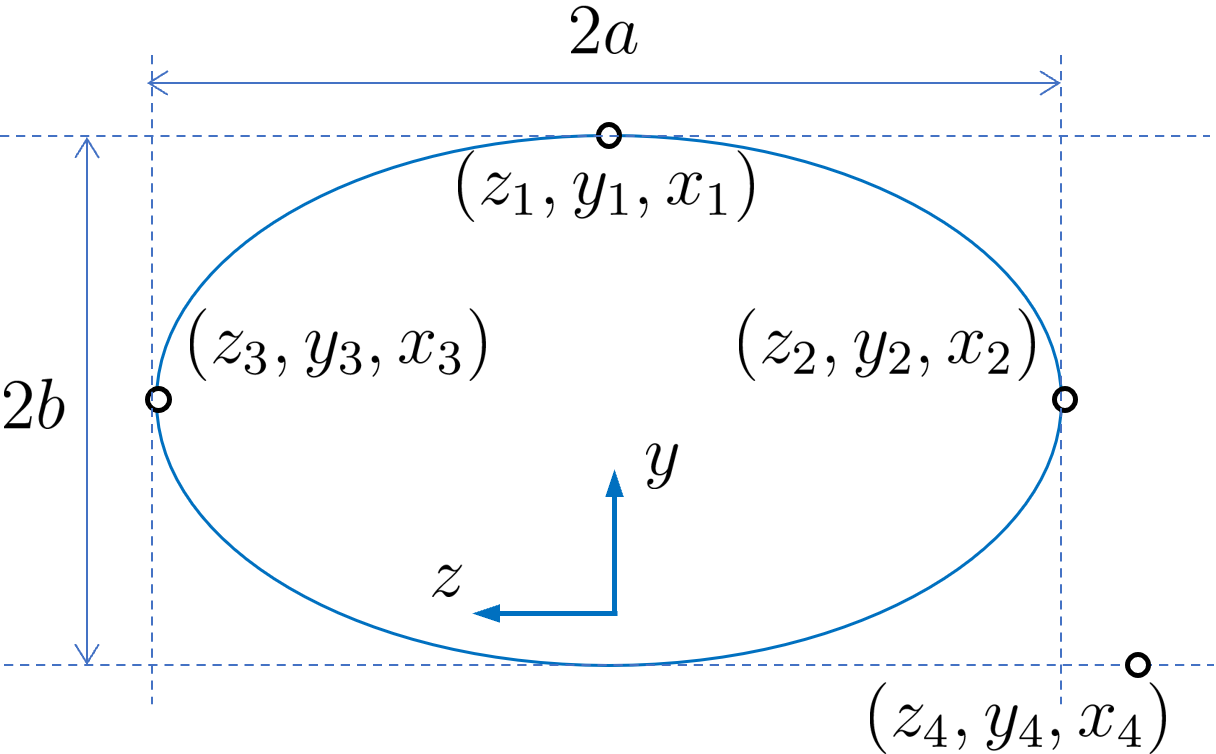}
 \caption{ZY (transverse) view of ellipsoid illustrating where points are collected to fit the ellipsoid.}
 \label{fig:ellipse}
\end{figure}

\begin{figure}[t]
 \centering
 \includegraphics[width=\columnwidth]{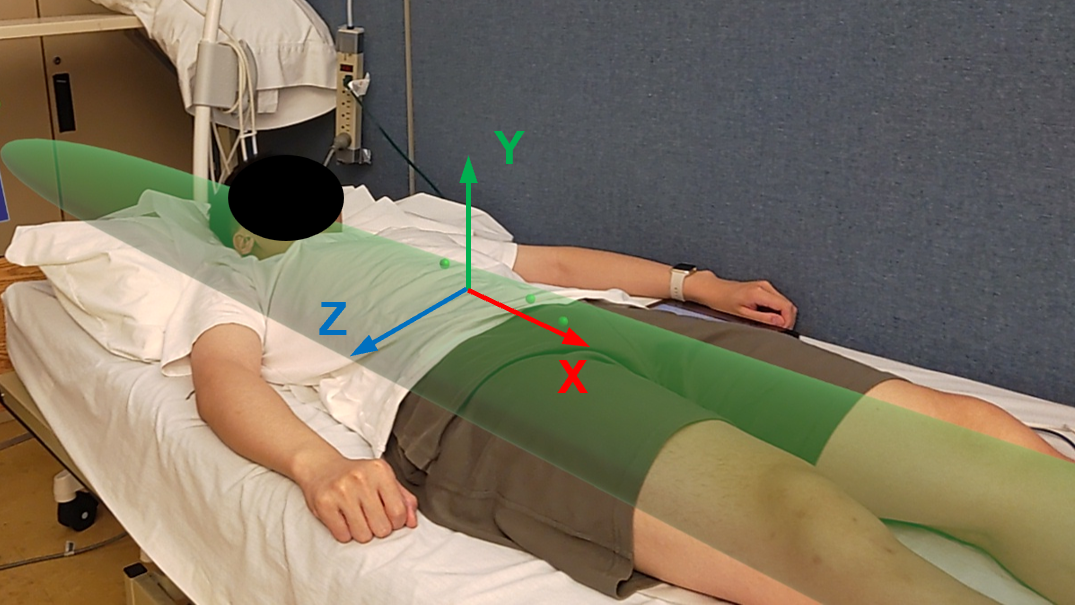}
 \caption{Fitted ellipsoid and coordinate system rendered over the patient's body.}
 \label{fig:ell_render}
\end{figure}

These points are illustrated in Fig. \ref{fig:ellipse} which shows a transverse view of the ellipsoid. The fourth point (``Bed level'') can be collected anywhere on the surface of the bed as only the height component ($y_4$) is used. The parameters can then be computed efficiently by the set of equations
\begin{align}
    a &= (z_3 - z_2) / 2 \\
    b &= (y_1 - y_4) / 2 \\
    z_c &= (z_3 + z_2) / 2 \\
    y_c &= (y_1 + y_4) / 2 \\
    x_c &= x_1
\end{align}
\noindent We can now specify the ellipsoid by
\begin{equation}
    \frac{(z - z_c)^2}{a^2} + \frac{(y - y_c)^2}{b^2} + \frac{(x - x_c)^2}{c^2} = 1
    \label{eq:ell}
\end{equation}

The resulting ellipsoid, rendered visually, is shown in Fig. \ref{fig:ell_render}. Using (\ref{eq:ell}) we determine when the expert's virtual transducer position ($\pmb{x}_e=[x_e,~y_e,~z_e]^\top$) penetrates the ellipsoid and render a force proportional to the penetration distance with a spring and damping constant and normal to the ellipsoid surface. To detect penetration into the ellipsoid, consider an ellipsoid that is uniformly scaled down from the original one such that $\pmb{x}_e$ lies on its surface. The surface normal vector at $\pmb{x}_e$ is then 
\begin{align}
    \pmb{n} &= \begin{bmatrix}
        \frac{(x_e - x_c)}{c^2}\\\frac{(y_e - y_c)}{b^2}\\\frac{(z_e - z_c)}{a^2}
    \end{bmatrix}\nonumber\\
    \hat{\pmb{n}} &= \pmb{n}/\|\pmb{n}\|
\end{align}

\noindent This is approximately equal to the surface normal vector at the closest point on the original ellipsoid. To find the penetration distance, $d$, we find where $\pmb{x}_e+d\hat{\pmb{n}}$ intersects the ellipsoid by solving the following quadratic:
\begin{equation}
    (\pmb{x}_e+d\hat{\pmb{n}}-\pmb{x}_c)^\top P^{-1} (\pmb{x}_e+d\hat{\pmb{n}}-\pmb{x}_c) = 1
\end{equation}

\noindent Where $P$ is a diagonal matrix with elements $a^2$, $b^2$, and $c^2$. The quadratic gives two solutions, the near and far-side intersections with the ellipsoid, so we choose the smaller of the two. If $d>0$, the haptic device is intersecting the ellipsoid and we apply a force 
\begin{equation}
    \pmb{f}=dK_p\hat{\pmb{n}} - K_d\pmb{v}
\end{equation}
Where $K_p$, $K_d$ are diagonal gain matrices and $\pmb{v}$ is the velocity.

\subsection{User Study}
Approval for all ethical and experimental procedures and protocols was granted by the University of British Columbia Clinical Research Ethics Board under application number H23-02587 on March 26, 2024. We performed a total of 11 abdominal US scans on healthy volunteers from the Skidegate community. Participants ($n=10$) of various ages (ranging from 20 to 60 years +) and backgrounds, and without prior experience using the human teleoperation system, were recruited as followers. One follower performed 2 scans and the rest performed 1 scan. Two certified sonographers with 30 + years of US experience took turns guiding the followers from the expert side to perform the scans. The followers and patients were situated in Skidegate, Haida Gwaii while the expert sonographers were situated 754 km away in Vancouver, Canada. The expert side system was connected to the Internet via Ethernet while the follower side system was connected via WiFi. For each scan, the sonographers were tasked with acquiring the following five target images and measurements:
\begin{enumerate}
    \item Proximal aorta with the anteroposterior (AP) diameter
    \item Longitudinal view of the proximal inferior vena cava (IVC)
    \item Long axis view of the left lobe of the liver
    \item Transverse view of the left lobe of the liver
    \item Transverse view of the right portal vein
\end{enumerate}

The two sonographers were involved with the development of the system and as such had several opportunities to gain familiarity with the system during the few months leading up to this study. Before each scan, the follower was given instructions through a presentation with short video demonstrations. The video conferencing call allowed the sonographer to ask the patients to hold their breath at various times during the scan as normally done in a conventional abdominal US exam.

After each scan, the follower completed a NASA TLX survey \cite{Hart:1988} and a modified system usability scale questionnaire. After all the scans, both sonographers filled out a questionnaire assessing their experience using the system. Image quality was assessed by two radiologists on a six point scale based on the image generation assessment tool introduced and validated in \cite{Millington:2016,Millington:2017}. Both of these studies used the same scale for image generation which is shown in Table~\ref{tab:imageScale}. This scale demonstrated good inter-rater reliability and was able to detect change in learner performance.
\begin{table}[t]
    \caption{Image quality scores and definitions \cite{Millington:2016,Millington:2017}. The radiologists provided a score for each individual image.}
    \label{tab:imageScale}
    \scriptsize
    \centering
    \begin{tabularx}{\columnwidth}{X c}
        \toprule
        Score Definition &  Score\\
        \midrule
        Not obtained & 0\\
        Image quality is too poor to permit meaningful interpretation & 1\\
        Image quality is better than poor but less than suboptimal for interpretation & 2\\
        Suboptimal image quality, but basic image interpretation is possible & 3\\
        Better than suboptimal but not good & 4\\
        Good image quality, meaningful image interpretation is easy & 5\\
        \bottomrule
    \end{tabularx}
\end{table}

The alignment error of the followers was also assessed by recording the relative positions and orientations of the physical and virtual transducers, determining the difference at each respective time step, and using the difference to compute the root mean squared error (RMSE) for each scan with respect to position and orientation. The position and orientation of the physical transducer was obtained using the IR pose tracking feature described in the \textit{System Design} subsection. A centered RMSE was also obtained for each scan by subtracting the mean error from each data point to isolate relative alignment precision.

\subsection{Data analysis}
Descriptive statistics such as the means and standard deviations of the image quality scores, follower usability scores, task load scores, RMSE and centered RMSE were determined. The distribution of these results were visualized to identify any patterns. The Spearman's rank correlation coefficient was computed between image quality score and RMSE to determine the relationship.

\section{Results}
\subsection{Image Quality}

\begin{figure}
 \centering
 \includegraphics[width=\columnwidth]{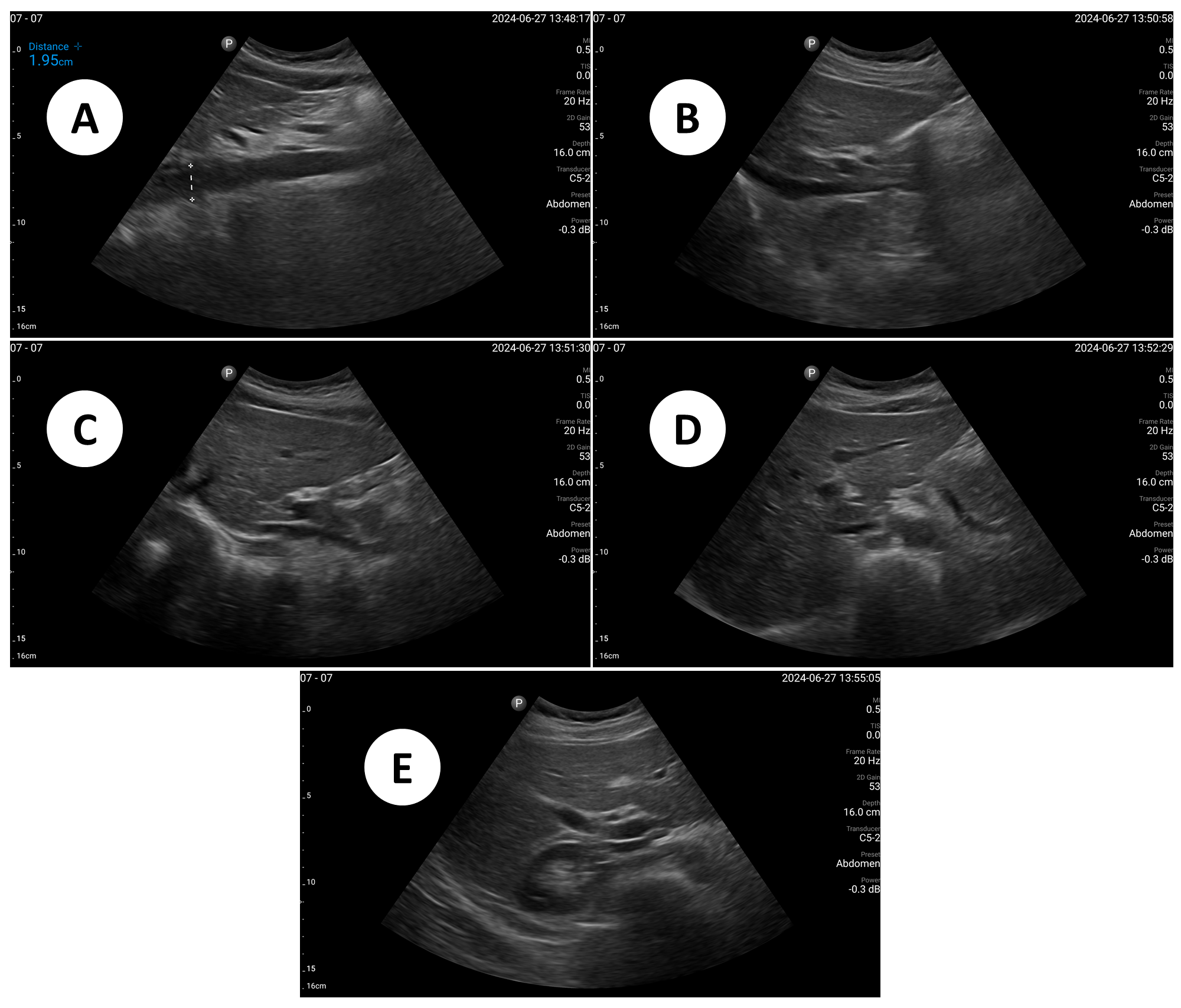}
 \caption{Example US images acquired in the study using the human teleoperation system. A: Proximal aorta (AP diameter). B: IVC - longitudinal. C: Left lobe of liver - long axis. D: Left lobe of liver - transverse. E: Right portal vein - transverse. }
 \label{fig:USimages}
\end{figure}

\begin{figure}[t]
 \centering
 \includegraphics[width=0.9\columnwidth]{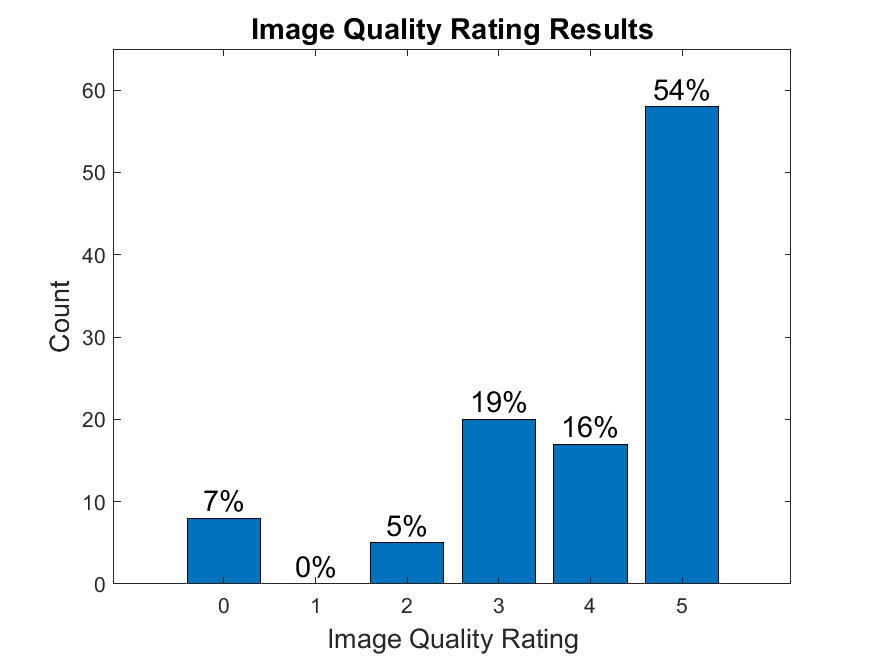}
 \caption{Image quality scores. The proportion of each score is indicated above each respective bar.}
 \label{fig:imageQualityBar}
\end{figure}

\begin{figure}[t]
 \centering
 \includegraphics[width=0.9\columnwidth]{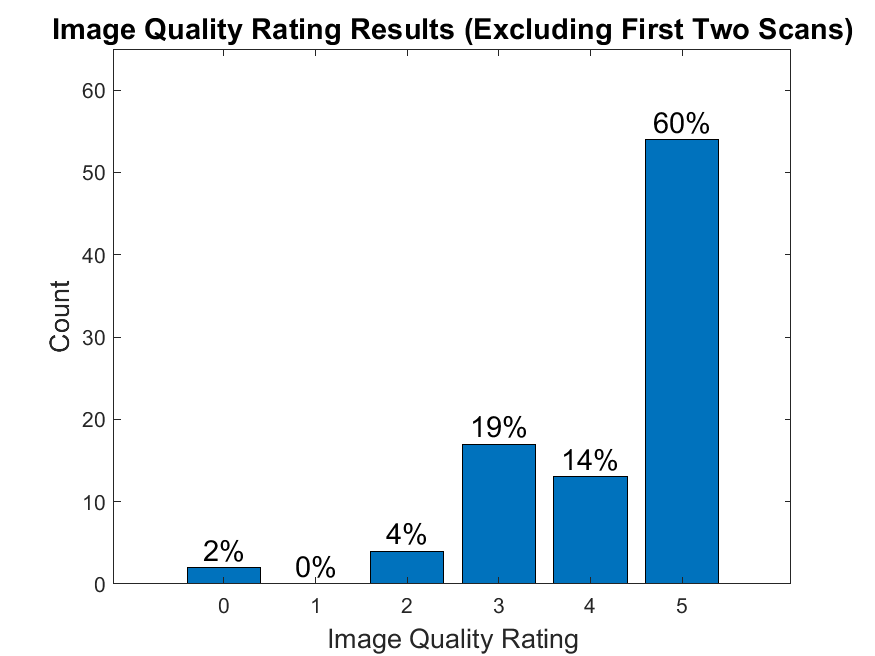}
 \caption{Image quality scores excluding the first two scans.}
 \label{fig:imageQualityBarRemoved}
\end{figure}

The sonographers completed 11 abdominal US scans, each with 5 target images and measurements for a total of 55 images. An example image of each of the targets acquired during these tests is shown in Fig. \ref{fig:USimages}. The image quality results are shown in Fig. \ref{fig:imageQualityBar} which illustrates the total number of each score given by both radiologists. The results excluding the first 2 scans are also shown in Fig. \ref{fig:imageQualityBarRemoved} because some technical issues such as internet disconnections occurred during these scans which may have affected the results. Of the 55 target images, 7 were considered missing by at least one of the radiologists. Of these missing targets, 3 were not captured due to large amounts of bowel gas and body habitus while one was seen but the sonographer did not capture and save the image. When considering only the images that were captured, a mean score of $4.28 \pm 0.95$ out of 5 was obtained and 91.7\% of the images were scored $\ge 3$ by both radiologists. 31.3\% of the images were scored 5 by both radiologists, indicating the image quality was good and meaningful image interpretation was easy.

\subsection{Usability and Task Load}
\begin{figure}[t]
 \centering
 \includegraphics[width=\columnwidth]{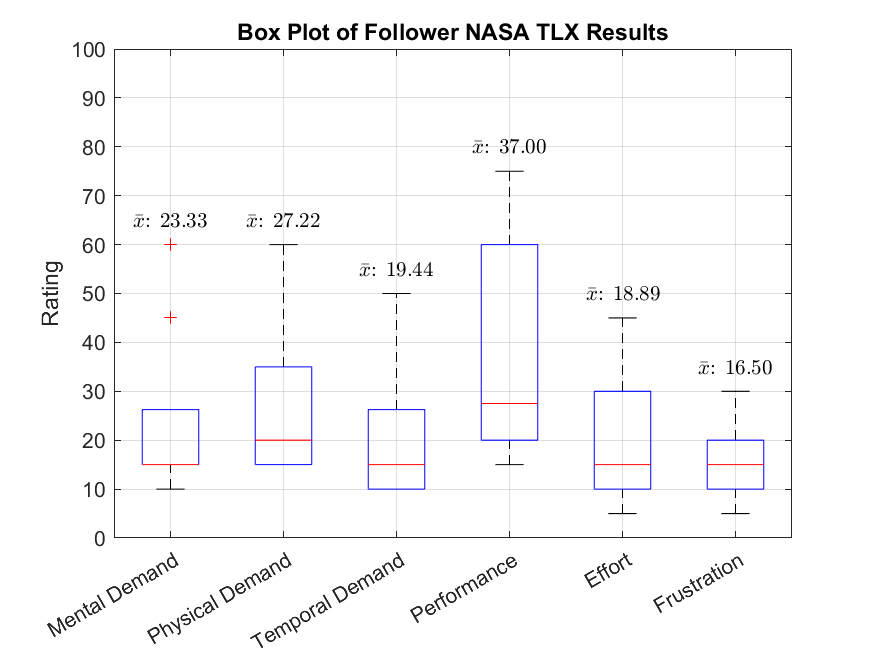}
 \caption{Boxplot and mean scores (out of 100; a lower rating is better, even for performance) from the NASA TLX survey completed by the followers ($n=10$).}
 \label{fig:nasatlx}
\end{figure}

Overall, the NASA TLX and system usability scale results showed low workload and high usability. The mental demand, effort, and frustration subscales had mean ratings of 23.3, 18.9 and 16.5 respectively out of 100, where a lower score indicated lower workload (Fig. \ref{fig:nasatlx}). The performance subscale had a mean rating of 37.0, where a lower score indicated better performance.

The follower usability questions and mean scores are shown in Table~\ref{tab:usability}. All 10 followers agreed or strongly agreed the system was easy to use. All 10 followers also agreed or strongly agreed most people would learn to use the system very quickly. In the questionnaire completed by the sonographers, both indicated the expert side of the system was easy to use and the control of the US transducer using the haptic device was intuitive. The mean completion time to acquire all target images and measurements was 645±116 seconds. The sonographer questionnaire results can be found in the appendix (section \ref{appendix:expque}).
\begin{table}[h]
    \caption{Mean scores from the modified system usability scale completed by the followers ($0=$ strongly disagree, $4=$ strongly agree).}
    \label{tab:usability}
    \scriptsize
    \centering
    \begin{tabularx}{\columnwidth}{X c}
        \toprule
        Usability Prompt/Question &  Mean $\pm$ SD\\
        \midrule
        I thought the overall system was easy to use & $3.6 \pm 0.52$\\
        I found it easy to follow the virtual transducer & $3.6 \pm 0.52$\\
        I found the point/dot fitting at the beginning easy to do & $3.7 \pm 0.48$\\
        I found it easy to follow the sonographer/expert’s instructions & $4.0 \pm 0$\\
        I needed to learn a lot of things before I could get going with the system (lower is better) & $1.4 \pm 0.97$\\
        I would imagine that most people would learn to use this system very quickly & $3.5 \pm 0.53$\\
        I found the system very cumbersome to use (lower is better) & $1.2 \pm 1.23$\\
        I felt very confident using the system & $3.4 \pm 0.70$\\
        \bottomrule
    \end{tabularx}
\end{table}

After completion of the scans, the followers were also asked to provide verbal feedback on their experiences. This offered additional insights into the human computer interaction aspects of this system not captured by the questionnaires. Many followers indicated it was easiest to follow the virtual transducer by looking at the transducer body itself rather than the markers on top. It was also often mentioned that the geometry and size of the transducer's force sensor shell made it difficult to grasp and rotate the transducer comfortably. All of the followers indicated a sense of trust towards the system and enthusiasm to use it again, suggesting there would likely be high acceptance of this technology; however, further investigation is required to validate this.

\subsection{Alignment Error}
\begin{figure}[t]
 \centering
 \includegraphics[width=\columnwidth]{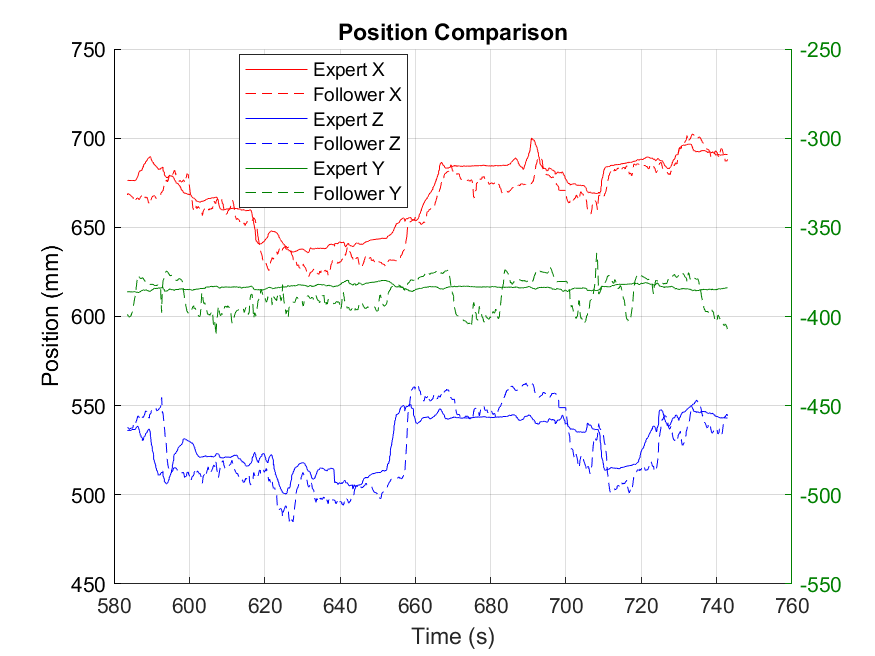}
 \caption{Example trajectory of follower's transducer (dotted lines) following the virtual transducer's trajectory (solid lines).}
 \label{fig:posErrSample}
\end{figure}

\begin{figure}[t]
 \centering
 \includegraphics[width=\columnwidth]{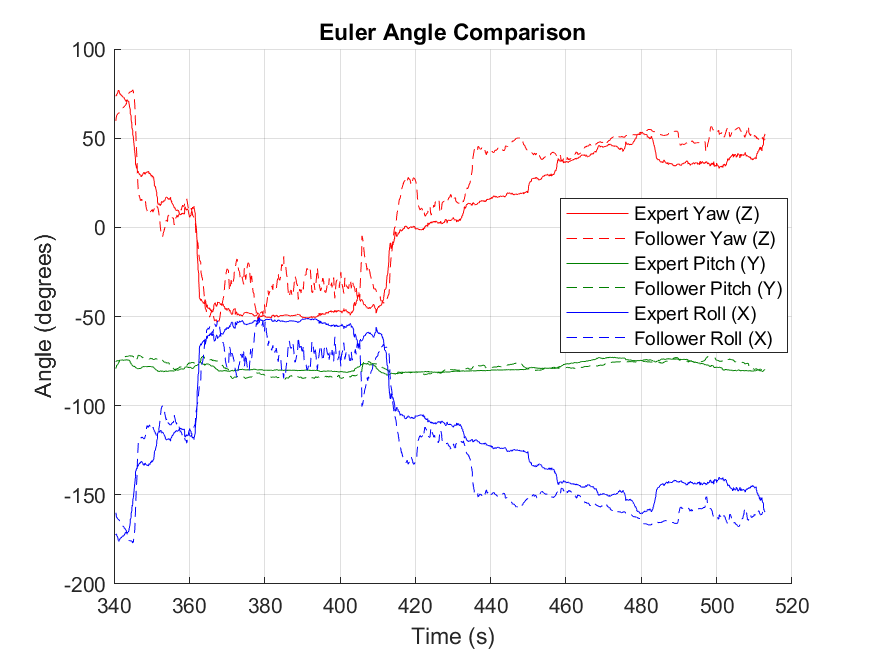}
 \caption{Example orientation trajectory of follower's transducer (dotted lines) following the virtual transducer's orientation trajectory (solid lines) represented as Euler angles.}
 \label{fig:quatErrSample}
\end{figure}

An example of the tracking data collected to compute alignment error is illustrated in Fig. \ref{fig:posErrSample} and Fig. \ref{fig:quatErrSample} and the RMSE values of each scan are presented in Table~\ref{tab:tracking}. Overall, the followers achieved a mean position RMSE of $31.94 \pm 10.17$ mm and mean orientation RMSE of $11.39 \pm 3.55 ^\circ$. The mean centered position and orientation RMSE was $10.06 \pm 3.74$ mm and $5.81 \pm 3.33 ^\circ$ respectively. No statistically significant correlation was observed between alignment error (both position and orientation) and image quality. The Spearman's rank correlation coefficient was 0.20 ($p=0.55$) and 0.00 ($p=1.00$) for position and orientation respectively.
\begin{table}[t]
    \caption{Position and orientation RMSE for each scan. The centered RMSE was computed from the error values with the mean error subtracted.}
    \label{tab:tracking}
    \scriptsize
    \centering
    \begin{tabularx}{\columnwidth}{X X X X X}
        \toprule
        Scan Number &  Position RMSE (mm) & Centered Position RMSE (mm) & Orientation RMSE ($^\circ$) & Centered Orientation RMSE ($^\circ$)\\
        \midrule
        1  &  31.9 & 14.2 & 18.3 & 5.70 \\
        2  &  41.3 & 6.99 & 11.5 & 3.82 \\
        3  &  29.8 & 8.88 & 8.04 & 3.73 \\
        4  &  30.8 & 8.09 & 9.21 & 3.72 \\
        5  &  44.3 & 18.5 & 15.2 & 4.19 \\
        6  &  32.7 & 11.3 & 11.5 & 10.9 \\
        7  &  22.7 & 9.13 & 6.46 & 2.63 \\
        8  &  51.8 & 11.6 & 15.1 & 13.2 \\
        9  &  19.3 & 5.14 & 8.89 & 5.16 \\
        10 &  22.7 & 8.90 & 11.5 & 6.94 \\
        11 &  24.0 & 7.84 & 9.59 & 3.96 \\
        \midrule
        $\text{Mean} \pm \text{SD}$ & 31.9 $\pm$ 10.2 & 10.1 $\pm$ 3.7 & 11.4 $\pm$ 3.55 & 5.81 $\pm$ 3.33 \\
        \bottomrule 
    \end{tabularx}
\end{table}

\section{Discussion}

Both radiologists gave a score $\ge 3$ to $91.7\%$ of the acquired images, indicating these images were unanimously considered of interpretable quality. 5 out of the 7 images where at least 1 of the radiologists gave a score of 0 (indicating missing target) occurred in the first 2 scans. These first 2 scans encountered technical issues such as internet disconnections which were resolved in later scans. Without the first two scans, there is a noticeable increase in the proportion of image quality scores $\ge 3$ as shown by comparing Fig. \ref{fig:imageQualityBar} and Fig. \ref{fig:imageQualityBarRemoved}. 

Comparing the image quality scores to the corresponding RMSE, there was no statistical evidence of a relationship between the two variables. Interestingly, the eighth scan which had the largest position RMSE and second largest orientation RMSE received the highest image quality score. The sonographer noted that this patient had an ideal body habitus, suggesting this may have a greater effect on image quality than alignment error.

The alignment RMSE values decreased significantly after centering as shown in Table~\ref{tab:tracking}. This suggests many of the followers had a constant offset in their alignment. A likely cause of this is the improper wearing of the HoloLens 2 headset, leading to a misalignment between the perceived and actual positions of the virtual transducer. In scan 8, the follower initially had a large RMSE of 51.8 mm which decreased to 11.6 mm after centering, indicating a significant constant offset. Despite this, the sonographer noted they did not notice any substantial inaccuracy during scanning. Moreover, all images acquired during this scan received the highest image quality score from both radiologists out of all the scans. A likely explanation for this is that the offset can be implicitly accounted for by the sonographer, who relies on the ultrasound image to infer the true position of the transducer. Thus, relative alignment precision rather than absolute accuracy is most important.

The NASA TLX ratings from our followers were compared to the reference values reported in \cite{Hertzum:2021}, which were based on the mean scores from 556 studies. These reference values encompassed various domains, including healthcare and virtual reality technologies. The mean values from our results were consistently lower (i.e. better) across all categories compared to the healthcare reference values. Our results also surpassed those for virtual reality technologies in all categories except physical demand, where the difference was a negligible one point. 

System usability results showed that complete novices from diverse backgrounds found the system easy to use as followers. All followers stated the initial brief instructions were sufficient and the sonographer’s verbal cues during scans were easy to follow. Many noted that when the sonographers verbalized their next move (e.g., rotating the transducer), it helped them anticipate and follow motions more easily, though this was not measured quantitatively. Some followers expressed uncertainty about their alignment with the virtual transducer, reflected in the relatively higher variance of the NASA TLX performance scores (Fig. \ref{fig:nasatlx}). Without real-time alignment feedback, followers may make unnecessary micro-adjustments, leading to less stable images. Future work will explore addressing this by providing alignment feedback, such as simple visual cues.

Feedback from the expert sonographers was mostly related to the control of the US imaging parameters. The Windows Phone Link app was used to allow the sonographers to remotely interact with the Philips Lumify Ultrasound app, but the sonographers found it difficult to use the cinescrolling feature, place calipers in the image and make annotations. In future work, we will create a more intuitive dedicated screen sharing and control app to address these issues, as started in \cite{Black:2024}. The sonographers also noted it would have helped to have direct feedback of the force applied by the follower. To address this, future work will explore other methods of force rendering such as position-force feedback or using a mesh of the patient acquired with the HoloLens 2, which would be more accurate and can be updated more easily compared to estimating the patient using an ellipsoid.

While this study demonstrated the feasibility of using human teleoperation for tele-US, additional scans in other abdominal regions would further validate the system's capabilities. Future work will involve evaluating human teleoperation for complete abdominal aorta examinations, where the sonographer will scan through the length of the aorta and capture measurements of its different segments. These quantitative measurements will be ideal for comparison with robotic or tele-mentored US. Further tests will explore the use of human teleoperation for renal ultrasound, which will require the patient to switch between supine and lateral decubitus positions, adding complexity.

The data collected through these tests will also present an opportunity to investigate integrating artificial intelligence (AI) for autonomous US guidance with mixed reality. The use of learning from demonstration and reinforcement learning for enabling autonomous US has previously been explored \cite{Burke:2023,Deng:2021,Li:2021,Droste:2020,Mylonas:2013}. By leveraging the rich, multi-modal data obtained with the human teleoperation system, it may be possible to train an AI agent to replace the expert or augment certain sub-tasks, a solution that would be especially valuable in areas with limited or no internet connectivity.

\section{Conclusion}
The study presented in this paper demonstrates the feasibility of using human teleoperation to perform remote abdominal US scans with a significant distance separating the sonographers and patients. Expert sonographers were able to acquire 48 out of the 55 possible target images across 11 scans. Of these acquired images, 92\% were considered sufficient quality for image interpretation by both radiologists who evaluated them. The 10 novice followers all expressed the system was easy to use, even with the minimal instructions provided. We found no correlation between the followers' alignment errors and the image quality scores given by the radiologists. Ultimately, these results show the potential impact human teleoperation can have for remote communities without local access to US imaging.

\addtolength{\textheight}{-3cm}   



\section*{APPENDIX}

\section{Expert Questionnaire Results}
\label{appendix:expque}
Table~\ref{tab:expertquestions} shows the responses to the usability questionnaire completed by the two sonographers.
\begin{table}[h]
    \caption{Questionnaire responses from the expert sonographers after all scans were completed. All scores are out of 5 where higher is better.}
    \label{tab:expertquestions}
    \scriptsize
    \centering
    \begin{tabularx}{\columnwidth}{>{\hsize=1.4\hsize\linewidth=\hsize}X >{\hsize=0.8\hsize\linewidth=\hsize}X >{\hsize=0.8\hsize\linewidth=\hsize}X}
        \toprule
        Question &  Sonographer 1 & Sonographer 2\\
        \midrule
        Was the system frustrating or easy to use? & $4$ & $4.5$\\
        Was it intuitive to move the haptic device based on the US image without your hand directly holding the probe on the patient & $5$ & $5$\\
       How much was hand motion limited by the haptic device & $5$ & $5$\\
        How natural did the force feedback feel & $3$ & $5$\\
       How effective was control of the US imaging parameters & $5$ & $5$\\
        How easy was control of the US imaging parameters & $3$ & $2.5$\\
       How good was the US image stream quality & $5$ & $5$\\
        How good was the US image stream delay & $4$ & $5$\\
        Was POV video stream important for rough positioning of the probe & Yes & Yes \\
        Was POV video stream useful to understand the patient and follower environment & Yes & Yes \\
       How was the POV video quality & Good & Good\\
        How was the POV video stream lag & $5$ & $5$ \\
        How much did you rely on the POV video stream vs US image & Mostly US but stream was helpful & Mostly US but stream was helpful \\
        \bottomrule
    \end{tabularx}
\end{table}

\section*{ACKNOWLEDGMENT}
The authors gratefully acknowledge funding support from NSERC and the Charles Laszlo Chair in Biomedical Engineering from The University of British Columbia, as well as infrastructure, technical, and funding support from Rogers Communications and MITACS, Canada.

\bibliographystyle{elsarticle-num} 
\bibliography{refs.bib}

\end{document}